**RESEARCH ARTICLE**

# Relativistic conceptions of trustworthiness: Implications for the trustworthy status of national identification systems

Paul R. Smart* , Wendy Hall and Michael Boniface

Electronics and Computer Science, University of Southampton, University Road, Southampton SO17 1BJ, United Kingdom
*Corresponding author. E-mail: ps02v@ecs.soton.ac.uk

**Abstract**

Trustworthiness is typically regarded as a desirable feature of national identification systems (NISs); but the variegated nature of the trustor communities associated with such systems makes it difficult to see how a single system could be equally trustworthy to all actual and potential trustors. This worry is accentuated by common theoretical accounts of trustworthiness. According to such accounts, trustworthiness is relativized to particular individuals and particular areas of activity, such that one can be trustworthy with regard to some individuals in respect of certain matters, but not trustworthy with regard to all trustors in respect of every matter. The present article challenges this relativistic approach to trustworthiness by outlining a new account of trustworthiness, dubbed the expectation-oriented account. This account allows for the possibility of an absolutist (or one-place) approach to trustworthiness. Such an account, we suggest, is the approach that best supports the effort to develop NISs. To be trustworthy, we suggest, is to minimize the error associated with trustor expectations in situations of social dependency (commonly referred to as trust situations), and to be trustworthy in an absolute sense is to assign equal value to all expectation-related errors in all trust situations. In addition to outlining the features of the expectation-oriented account, we describe some of the implications of this account for the design, development, and management of trustworthy NISs.

**Policy Significance Statement**

To be trustworthy, national identification systems (NISs) should strive to ensure that trustor expectations are accurate, in the sense that they are aligned with the capabilities of a NIS. This calls for open and honest communication about the properties of a NIS. Trustworthy NISs should also be bound by a set of commitments (or "policies") that serve as the basis for trustor expectations, while also acting as constraints on system behavior. This calls for greater attention to technologies that support policy-aware modes of operation. Finally, the emphasis on error minimization directs attention to a range of cognitive and communicative abilities that are intended to mitigate the possibility of misplaced trust. This speaks to a potential role for artificial intelligence systems in the design of trustworthy NISs.

> But he that is a friend to all men, is a friend to no man, and least of all to himself. For he must promise so much, that he cannot perform withall: and so breaking promise with some, he is trusted at length by none.
>
> Crook (1658)

## 1. Introduction

The term "Trustworthy National Identification System" suggests that trustworthiness is a particular feature of a national identification system (NIS),[1] one that can be evaluated independently of the specific individuals (i.e., trustors) who might be required to place their trust in the NIS. This is what might be called an absolutist (or one-place) view of trustworthiness. The absolutist view allows for the idea that trustees (e.g., a specific NIS) can be trustworthy in a highly generic sense, which is to say that a given trustee can be equally trustworthy to all actual and potential trustors across a range of trust-related situations.

Perhaps unsurprisingly, the absolutist view of trustworthiness has not gone unchallenged. For the most part, trust theorists have favored a more relativistic conception of trustworthiness. According to this relativistic conception, trustworthiness is not invariant across trust-related contexts. Instead, trustworthiness is deemed to vary according to both the trustor and the matters with which the trustee is trusted. This makes intuitive sense, since, in the case of human trustees, it seems highly unlikely that a given individual could be said to be trustworthy to absolutely everyone. Rather than being trustworthy to everyone, human trustees are selectively trustworthy—they are trustworthy to some individuals, but not to other individuals. In addition, human trustees are not trustworthy with regard to every matter with which they might be trusted. Someone may be highly trustworthy when it comes to the provision of financial advice, for example, but they may not trustworthy when it comes to looking after one's children (see Hardin, 2001).

These intuitions serve as the basis for a number of theoretical accounts of trustworthiness, all of which adopt a relativistic view of trustworthiness. These include so-called two-place accounts of trustworthiness, where trustworthiness is relativized to particular trustors, and three-place accounts of trustworthiness, where trustworthiness is relativized to particular trustors and particular spheres of activity (or domains of interaction) (for recent reviews, see Vivian, 2019; Carter and Simion, 2020). Both two-place and three-place accounts of trustworthiness reject the idea that we can understand trustworthiness as an intrinsic property of a trustee. This, it should be clear, has implications for the way we think about trustworthiness, as well as the way we develop and evaluate trustworthy systems. In particular, a relativistic conception suggests that it is a mistake to evaluate the trustworthiness of a system independently of the specific circumstances in which the system will be used.

Relative to the contrast between relativistic and absolutist views, it is arguably the absolutist view that best serves the interests of those involved in the effort to deliver trustworthy NISs. This is because the very notion of a trustworthy NIS implies that the relevant NIS has the capacity to be trustworthy in a highly generic sense, for example, to all the citizens of a nation state. This idea is challenged by theoretical accounts that countenance a relativistic approach to trustworthiness. In particular, a relativistic view seems to question the basic possibility of trustworthy systems that are deployed at a national or supranational level. The reason for thinking this is that it is hard to see how a single trustee could be equally trustworthy to trustors with competing or conflicting interests. This problem is arguably exacerbated by the presence of multiple stakeholder groups (users, developers, politicians, commercial organizations, and other social actors) who may have different vested interests regarding the functional operation of a NIS.

In the present article, we suggest that the interests of the NIS community are best served by an absolutist view of trustworthiness. While this is counter to the bulk of the theoretical literature—much of which is wedded to a relativistic conception of trustworthiness—we suggest that it is possible for NISs to function in a manner that is consistent with an absolutist view of trustworthiness. As a means of providing theoretical support for this idea, we outline a new account of trustworthiness, dubbed the expectation-oriented account. One of the virtues of this account is that it allows for the possibility of an absolutist view of trustworthiness. A second virtue is that it is readily applicable to nonhuman entities, such as NISs.

---

[1] For the purposes of the present article, the term "national identification system" refers to what is sometimes called a foundational identification system, that is, a system that is "created to manage identity information for the general population and provide credentials that serve as proof of identity for a wide variety of public and private sector transactions and services" (World Bank Group, 2019, p. 217). We are, in addition, concerned with identification systems that qualify as digital identification systems, that is, systems that use digital technology throughout the identity lifecycle.

The structure of the article is as follows: In Section 2, we outline the case for a relativistic approach to trustworthiness and discuss some of the reasons why this approach might be problematic for claims regarding the trustworthy status of NISs. In Section 3, we challenge the arguments that motivate appeals to both two- and three-place trustworthiness. We also provide a brief introduction to the aforementioned expectation-oriented account of trustworthiness. Finally, in Section 4, we discuss some of the policy-related implications of the expectation-oriented account, especially as these implications relate to the design, development, and management of trustworthy NISs.

## 2. Trustworthiness and the Relativity Problem

Trust and trustworthiness are typically seen to be important to the success of NISs. The World Bank Group (2019), for example, suggests that:

> For an ID system to be successful, the population—including vulnerable groups—must have trust and confidence in its design and implementation, particularly with respect to the collection, use and protection of sensitive personal data (World Bank Group, 2019, p. 81).

Note that the emphasis here is on the *trust* concept. The goal is thus to ensure that NISs are trusted by the populace, which is to say that users have an attitude of trust toward the NIS.

There is, however, a problem with this: While trust is no doubt important to the success of a NIS, it cannot be the thing that motivates the design, development, and management of *trustworthy* NISs. The reason for this is that there is an important difference between trust and trustworthiness. While trust is typically conceptualized as an attitude that a trustor adopts toward a trustee—as in X (trustor) believes that Y (trustee) is trustworthy—trustworthiness is more typically seen as a property of the trustee—as in Y is trustworthy (see McLeod, 2020). This way of thinking about trust and trustworthiness challenges the primacy of the trust concept when it comes to NIS-related research. Consider, for example, that someone can be trustworthy, even if no one believes them to be trustworthy. Conversely, trustors can be the unfortunate victims of what is called misplaced trust—a state-of-affairs in which a trustor places their trust in a trustee (assuming the trustee is trustworthy) only to discover that the trustee is not, in fact, trustworthy. Suffice to say, misplaced trust cannot be the ultimate goal of those who seek to ensure the success of NISs, for there is no point in trying to cultivate trust in a trustee if the trustee should turn out to be untrustworthy. The goal, then, is not so much to ensure the trusted status of NISs; it is more to ensure that NISs are actually trustworthy. This shift in emphasis—from trust to trustworthiness—is one that is reflected in a number of recent research efforts (Beduschi, 2021; Maple et al., 2021).[2]

So, what does it mean for a NIS to count as trustworthy? The answer is, unfortunately, unclear. One reason for the lack of clarity relates to the absence of any sort of consensus regarding the foundational notion of trustworthiness. While there have been a number of attempts to develop a theoretical account of trustworthiness, it remains unclear which (if any) of these accounts is the right one to adopt.[3] What is more, the majority of efforts to analyze the concept of trustworthiness have limited their attention to the realm of human trustworthiness. This can be a source of problems when it comes to the attempt to apply theoretical accounts of trustworthiness to nonhuman (e.g., technological) trustees.

In a technological context, trustworthiness is typically understood with respect to the possession of certain properties or features. A nice example of this stems from the work of Maple et al. (2021). They suggest that trustworthy NISs are those that possess a number of features related to security, resilience, privacy, robustness, ethics, and reliability. This sort of approach resembles that adopted for trustworthy artificial intelligence (AI) systems, where the possession of certain features (e.g., safety, reliability, and fairness) is deemed to distinguish trustworthy systems from those of the (presumably) untrustworthy kind. As noted by Simion and Kelp (forthcoming), one of the drawbacks of these sorts of approaches—which are sometimes referred to as objective list theories—is that they fail to tell us why or how the identified features relate to trustworthiness. This does not mean that the features identified by Maple et al. (2021) are incorrect, it simply means that an account of trustworthiness ought to tell us why these features are important when it comes to judgments of trustworthy status.

---

[2] The contrast between trust and trustworthiness is echoed by the distinction between perceived trustworthiness and objective trustworthiness (see Castelfranchi and Falcone, 2020). In the present article, we are primarily concerned with *objective trustworthiness*. That is to say, we are concerned with the actual (objective) status of trustees as either trustworthy or untrustworthy.

[3] Theoretical accounts of trustworthiness include the encapsulated interest account (Hardin, 2002), the trust responsiveness account (Pettit, 1995), the Confucian account (Sung, 2020), the virtue account (Potter, 2002), the counting-on account (Jones, 2012), the goodwill account (Baier, 1986), the commitment account (Hawley, 2019), and the obligation account (Kelp and Simion, 2020).

At the most general level, trustworthiness appears to be a disposition to do certain things. In particular, it appears to be a disposition to do what one is trusted to do. If I trust you to deliver a parcel by 8 pm, then you are trustworthy to the extent that you have a disposition to deliver the parcel by 8 pm. This does not mean that you will succeed in your delivery-related efforts, for there may be forces and factors that prevent you from acting in the way you are disposed to behave. In this case, we would say that your disposition has been "masked" (see Johnston, 1992). Your trustworthiness inheres in your disposition to do what you are trusted to do, but outside forces and factors may prevent you from manifesting this disposition. That is to say, outside forces and factors may prevent you from manifesting your trustworthiness.

Perhaps, then, trustworthiness can be understood as a disposition to do what one is trusted to do. Schematically, if X (trustor) trusts Y (trustee) to $\Phi$, where $\Phi$ corresponds to whatever it is that Y is trusted to do, then Y will count as trustworthy to the extent that they are disposed to $\Phi$. If this is the case, then trustworthiness would correspond to a dispositional property, specifically one that is possessed by Y; X's belief that Y possessed this property would correspond to what we call trust in its attitudinal form,[4] and the act of placing trust (i.e., trust in its action form) would serve as the trigger for the manifestation of Y's trustworthiness.[5]

Despite its appeal, there are reasons to think that this way of thinking about trustworthiness cannot be correct. If trustworthiness were simply a matter of doing what we were trusted to do, then our trustworthiness would be called into question every time we were trusted to do something that we could not do. And what about situations where we are trusted to do things that we ought not to do? Suppose a terrorist trusts a driverless car to drive into a crowd of pedestrians. Does the car manifest its trustworthiness by doing what the terrorist wants it to do? What about if the car agreed to do what the terrorist trusted it to do, but then responded by transporting the terrorist to the local police station? Does this make the car trustworthy or untrustworthy?

The answer, it seems, depends on whose perspective we are adopting. Inasmuch as the car drives into the pedestrians, then it looks to be trustworthy to the terrorist but not the pedestrians. Conversely, if the car drives to the police station, then it looks to be trustworthy to the pedestrians but not the terrorist. But if we are to take these intuitions at face value, then it seems that trustworthiness *cannot* be a property of the trustee after all. For if trustworthiness were to be a genuine property of the trustee, then we would expect the trustworthiness of the car to remain the same regardless of the way we looked at each of the scenarios. This does not appear to be the case, however. Rather than trustworthiness being a property of the car, it seems as though trustworthiness is more in the "eye of the beholder"—the car will either be trustworthy to the terrorist or the pedestrians, but not both at the same time. If this is true, however, then we have a significant problem, for it seems that there is no way to evaluate the trustworthiness of the car in the absence of a consideration of whose perspective we are going to adopt during the course of the evaluative effort. In particular, it does not seem appropriate to say that the car is trustworthy in an absolute or simpliciter sense, for the statement "the car (Y) is trustworthy" leaves us none the wiser as to whom the car is trustworthy to. The best we can do, it seems, is say that the car is selectively trustworthy: it is trustworthy to some individuals but not to other individuals. Such forms of discrimination may be perfectly permissible when it comes to the distinction between terrorists and pedestrians, but the social cake can always be divided along less palatable lines (e.g., rich vs. poor, liberal vs. conservative, and vaccinated vs. unvaccinated), and this is a source of worry when it comes to systems that are intended to serve the interests of large and heterogeneous trustor communities.

---

[4] Note that there is a difference between trust-as-attitude and trust-as-action (see Hardin, 2001). In saying that X believes that Y is trustworthy, we are referring to trust as a form of attitude (the attitudinal form). Conversely, when we talk of X placing their trust in Y, we are referring to trust as a form of action. Within the trust literature, trust actions are sometimes referred to as "trusting behavior" or "behaviourally exhibited trust" (Bauer and Freitag, 2018).

[5] See McKitrick (2018), for a general discussion of trigger conditions for dispositional properties.

At this point, we encounter an important contrast between what we will call absolutist and relativistic views of trustworthiness. According to an absolutist view of trustworthiness, trustworthiness is an intrinsic property of a trustee—it is something that can be measured and observed independently of the interests, needs, and concerns of a given trustor community. This is what might be called one-place trustworthiness, since we do not need to qualify our claims about Y's trustworthiness. Schematically, we can say that Y is either trustworthy or untrustworthy.

It will probably come as no surprise to learn that the absolutist view has received little in the way of support from trust theorists. Most theoretical accounts of trustworthiness adopt a more relativistic (or relational) approach to trustworthiness, which is to say that they reject the idea that one can be uniformly trustworthy to all actual and potential trustors. This makes sense given what we said about the driverless car scenario. The driverless car, it seems, cannot be equally trustworthy to both a terrorist and a pedestrian at the same time; it must be trustworthy to either one or the other. This, however, seems to challenge the idea that trustworthiness can be viewed as an intrinsic property of a trustee—something that can be measured or observed independently of a trustor-specific perspective. Rather than being trustworthy in an absolute or simpliciter sense, it seems that trustworthiness must be relativized to particular trustors or trustor communities. In this sense, the expression "trustworthy NIS" is, at best, elliptical. It cannot mean that a NIS is trustworthy in an absolute or simpliciter sense; rather, it must mean that a NIS is trustworthy relative to a specific, albeit perhaps unspecified, community of trustors.

Relativistic views of trustworthiness can typically be divided into what are called two-place and three-place accounts of trustworthiness. According to two-place accounts, trustworthiness is relativized to particular trustors (or trustor communities). To help us understand the motivation for a two-place account, consider the two cases illustrated in Figure 1. Figure 1a depicts a state-of-affairs in which the trustee is a military drone that is affiliated to one side of an armed conflict (i.e., blue force). As indicated by the arrows in this diagram, the drone is deemed to be trustworthy to the members of blue force, but not to the members of red force. Accordingly, the drone cannot be trustworthy in an absolute sense; rather, it must be trustworthy in a relativistic sense: it is trustworthy relative to blue force but not trustworthy relative to red force.

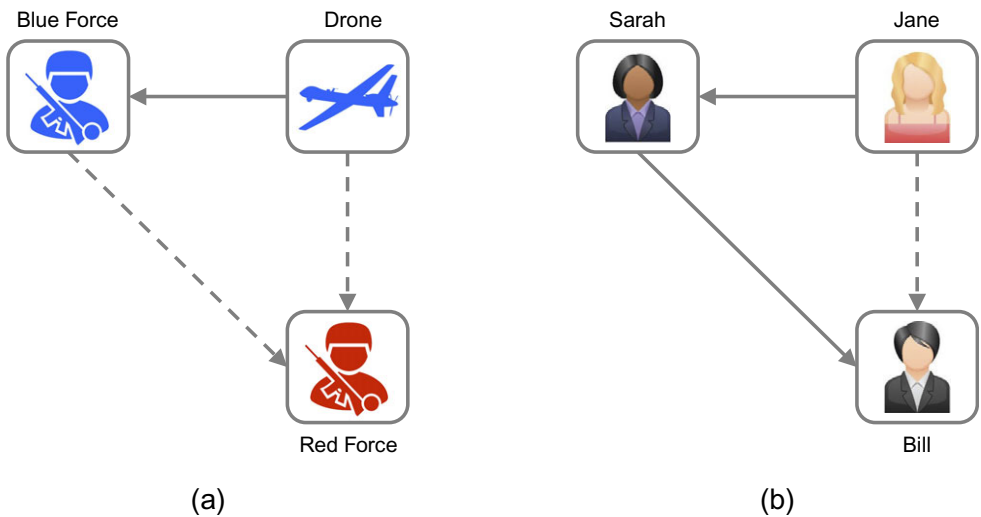

*Figure 1.* The relativization of trustworthiness to specific trustors. Arrows represent the trustworthiness/ untrustworthiness of the trustee (source) relative to the trustor (target). Solid lines symbolize trustworthiness, while broken lines symbolize untrustworthiness. In (a), a military drone is trustworthy to the members of blue force (solid line), but not to the members of red force (broken line). Similarly, in the case of (b), Jane is trustworthy to Sarah (solid line), but not trustworthy to Bill (broken line). Also note that in (a) a blue force soldier is not trustworthy to a red force soldier, whereas in (b) Sarah is trustworthy to Bill.

A similar sort of situation is depicted in Figure 1b. In this case, Jane is assumed to be the trustee. Jane is trustworthy to Sarah, but not to Bill. Whenever Sarah places her trust in Jane, Jane will respond by doing what she is trusted to do. Bill, by contrast, is much less fortunate. Any trust he places in Jane will be misplaced, for Jane will not respond by fulfilling his trust. Once again, we appear to encounter a state-of-affairs in which an absolutist view of trustworthiness looks to be utterly implausible. What is worrying about this case is that it seems to reflect the reality of human trust relationships. While we might pride ourselves in being trustworthy to a few close friends and family members, we would surely resist the temptation to say that we were trustworthy to absolutely everyone. And even if we tried to be trustworthy to everyone, it seems unlikely that we would succeed in this effort. Some trustors would, no doubt, end up being disappointed, and, in the worst case, we might end up being trusted by no one at all (recall the opening epigraph to the present article).

Three-place trustworthiness extends the relativization introduced by two-place trustworthiness. The idea here is that trustees are trustworthy to particular trustees with regard to particular matters. Schematically, Y is trustworthy relative to X in respect of $\Phi$. Again, this makes intuitive sense, for it seems unlikely that any given trustee could be trustworthy with respect to absolutely everything. A dentist, for example, might be trustworthy when it comes to the performance of certain dental procedures, but not trustworthy when it comes to the keeping of personal secrets. In this case, it would make no sense to say that the dentist is either trustworthy or untrustworthy; rather, the dentist's trustworthiness is scoped to particular matters or—as Jones (2012) suggests—domains of interaction.

The intuition behind three-place trustworthiness is nicely summarized in the following quotation from Robbins (2016):

> I may […] trust my wife, but not surely for anything and everything. I might trust her to edit a paper or maintain fidelity but not with medical advice or to fly a plane. Under these conditions, I might assume that her motivations toward my interests are the same, regardless of the matter at hand, but her ability to actualize each of these matters will vary. As a result, my beliefs about her trustworthiness will vary from matter to matter (Robbins, 2016, p. 978).

Robbins's wife, we may suppose, is favorably inclined toward her husband—she wants what is best for him. But these feelings of goodwill do not seem sufficient for trustworthiness. In addition to goodwill, it seems that Robbins' wife must possess certain abilities or competences that enable her to do what she is trusted to do. While this is broadly compatible with approaches that highlight the importance of abilities to trustworthiness (e.g., Mayer et al., 1995), it also raises a worry about the limits of trustworthiness/untrustworthiness. Quite plausibly, there are a finite number of things that we can do, but an infinite number of things that we cannot do. Does this mean that we are infinitely untrustworthy? And, if that is the case, then how could our untrustworthiness possibly be reduced?

A consideration of both two- and three-place trustworthiness yields a potential problem for those who seek to ensure the trustworthy status of NISs. The problem is that it does not seem appropriate to regard NISs (or, indeed, any system) as being trustworthy or untrustworthy in an absolute or simpliciter sense: we can only evaluate claims of trustworthiness relative to particular communities of trustors and particular domains of interaction. In one sense, this does not appear particularly problematic, for we could qualify our claims by stating that a NIS (Y) is trustworthy relative to the citizens of a particular nation state (X), but only in respect to certain matters ($\Phi$). The problem is that the citizens of a nation state can be a pretty variegated bunch, and it is far from clear that a single system can be designed so as to act in the interests of all citizens at the same time. What is more, there are typically a number of stakeholders involved in the effort to implement NISs (see World Bank Group, 2019, pp. 21–22), and these stakeholders may have different views as to what they want a NIS to do. As noted by Beduschi (2021, p. 4), "[t]he digital identity landscape includes a multitude of actors with different and often divergent interests, defining how technologies are designed and implemented." Given that inclusivity is one of the driving features of NIS-related research, it seems NISs ought to be trustworthy to all those who are apt to be affected by the system. But the question is whether this is really possible, both in practice and in principle. Inasmuch as we buy into the idea that trustworthiness ought to be conceptualized along relativistic lines, then the answer is likely to be a resounding "no," especially in situations where stakeholders have competing or conflicting interests. A NIS cannot be the trusty servant of multiple masters with competing interests, just as a military drone cannot be equally trustworthy to the opposing sides of an armed conflict. NISs, by their very nature, are intended to service the interests of a large (and typically heterogeneous) community of trustors. But insofar as we embrace a relativistic approach to trustworthiness, then trustworthiness becomes progressively harder to obtain the more we expand the scope of the X and $\Phi$ parameters. This is the essence of what we will call the *relativity problem*.

## 3. Tackling the Relativity Problem

Relative to the contrast between relativistic and absolutist views of trustworthiness, we suggest that it is the absolutist view that best serves the interests of those involved in the practical effort to deliver trustworthy NISs. Unfortunately, as we saw in the previous section, there are reasons to think that the absolutist view is the wrong way to think about trustworthiness. This is reflected in the majority of theoretical accounts of trustworthiness, many of which are wedded to a two- or three-place account of trustworthiness.

In the present section, we aim to outline an alternative account of trustworthiness, one that is compatible with *both* absolutist and relativistic views of trustworthiness.[6] As a means of introducing this account, it will be useful to temporarily suspend our interest in trustworthiness and consider the nature of promissory relations. In promissory relations, one individual (the promisor) makes a promise to another individual (the promisee), and the promise relates to things that the promisee will do (or refrain from doing). (Schematically, Y promises X to $\Phi/{\sim}\Phi$.) In making a promise, the promisee incurs a commitment or obligation to act in a manner that is consistent with the promise.

Given this basic characterization of promissory relations, we are now in a position to ask ourselves the following question: What is it that makes someone a *good* promisor? The answer, it seems, must have something to do with the way that a promise tells us something about the future behavior of the promisor. Good promisors thus have a robust tendency (or disposition) to keep their promises; bad promisors, by contrast, do not.

But is that all there is to being a good promisor? One reason to think this might not be the case stems from a consideration of the role played by promisors in the acquisition of promissory obligations. Suppose you aspire to be a good promisor. One way you can do this is by keeping the promises you make. But another (equally important) strategy is to avoid making promises that you know you cannot keep. One is not a good promisor if one makes promises that one is ill-disposed to keep, and this is so even if the promises one makes are ones that are actually kept. Consider, for example, that one is not a good promisor if one promises to repay a loan that cannot be repaid, and this remains the case even if one should unexpectedly win the lottery and thus be in a position to make good on the earlier promise. Good promisors do not just keep their promises, they also limit their promises to those that are keepable. What is more, good promisors do not just keep their promises as a result of some fortunate happenstance (such as winning the lottery)—they expect *themselves* to be in a position to fulfill their promises at the time the promises are made.

Such considerations are useful in helping us formulate a response to the relativity problem. Consider one of the motivations for two-place trustworthiness, namely, the idea that a single trustee cannot be equally trustworthy to multiple trustors with competing interests. According to the example given in [Figure 1](#)a, for example, a military drone cannot be equally trustworthy to the opposing sides of an armed conflict. This seems eminently plausible given the nature of armed conflicts. In such situations, one cannot be trustworthy to one's own forces and the forces of an opposing side at the same time. Indeed, it seems likely that part of what it means to be trustworthy to one side is to be untrustworthy to the opposing side. Accordingly, the drone cannot be trustworthy in an absolute sense; instead, it must be trustworthy in a relativistic sense.

---

[6] Note that we are not suggesting that two- and three-place views of trustworthiness are wrong, or that they have no place in our understanding of trust relations. We are merely attempting to allow for the possibility of an absolutist (or one-place) approach to trustworthiness. This is not, we suggest, a zero-sum game; we can allow for the possibility of both relativistic and absolutist views of trustworthiness.

All this, however, assumes that the trustworthiness of a trustee is called into question whenever the trustee acts in a manner that conflicts with the interests of a potential trustor. But is that really the case? The drone, we may assume, has never given the opposing side any reason to believe that it would fulfill the trust that members of the opposing side might place in it. This is important, for if no one (including red force) expects the drone to fulfill red force's trust, then it is unclear whether the drone could be accused of violating that trust. If red force knows that the drone will never act in their interests, then they will refrain from placing their trust in the drone. But if red force refrain from placing their trust in the drone, then it is hard to see how the drone could be accused of betraying (or failing to fulfill) this trust. If the drone never has an opportunity to violate red force's trust, and it always acts so as to fulfill blue force's trust, then is there any reason to regard it as untrustworthy?

At this point, the critic will no doubt want to insist that no member of blue force, technological or otherwise, could possibly be regarded as trustworthy to the members of red force: In order for the drone to be trustworthy to blue force, it has to act in a manner that is consistent with the interests of blue force. But part of what it means to act in the interests of blue force is to act in a manner that is contrary to the interests of red force. Given this, how could the drone possibly be trustworthy to both forces at the same time?

As a means of challenging this intuition, let us return to the realm of promissory obligations. Suppose we were tasked with evaluating the status of the drone as a good promisor. In order to make the case a little more compelling, let us assume the drone is equipped with a suite of cognitive and communicative capabilities, such that it is able to communicate promises and coordinate its behavior with respect to these promises. To be a good promisor, then, the drone must keep whatever promises it has made to blue force, many of which, we may suppose, are reflected in the algorithms that govern the drone's overt behaviour. This is all well and good, but it is hard to see why the drone's status as a good promisor would diminish in the eyes of red force. As long as the drone does not make any promises to red force, then it cannot be accused of breaking those promises. To be a good promisor, the drone must keep its promises. But it cannot count as a bad promisor simply because it avoids making promises that cannot be kept.

Much the same, we suggest, applies to the drone's status as a trustworthy entity. In order to be trustworthy, the drone must act in accordance with the expectations of blue force, and some of these expectations pertain to the way the drone behaves toward the members of red force. But in fulfilling the trust of blue force, the drone is not required to act in a manner that violates the trust of red force. The reason for this is simple: red force members are not being given an opportunity to place their trust in the drone, and if such an opportunity were to arise, the drone would (let us suppose) respond in such a way as to make it clear that it could not fulfill the trust that red force were attempting to place in it. This does not seem incompatible with the drone's status as a trustworthy entity. Indeed, if we were to modify the case, such that the drone was substituted with a human soldier, then we would not say that the soldier was untrustworthy simply because they refused to defect to an opposing side. Nor is it clear that such an individual would be regarded as untrustworthy by the members of the opposing side! To be sure, the opposing side would not be able to rely on the soldier for anything. But, at the same time, there is no sense in which the opposing side is being misled into thinking that the soldier will do something that the soldier subsequently does not do. Inasmuch as the soldier is trustworthy, then they will simply refuse to be bound by whatever trust-related obligations red force seeks to impose upon them. There is, we suggest, no sense in which the soldier could be accused of betraying anyone's trust in this situation, and if there is no possibility of betraying anyone's trust, then it is unclear why anyone ought to regard the soldier as untrustworthy.

In Figure 1b, we suggested that Jane is trustworthy relative to Sarah, but that she is not trustworthy relative to Bill. This suggests that Jane must be trustworthy in a relativistic sense. Indeed, given the way the case is presented, it should be clear that the relativistic conclusion is guaranteed: If Jane is indeed trustworthy to Sarah but not to Bill, then she cannot be said to be trustworthy in an absolute sense.

Suppose, however, that we add some extra details to the case. What if Jane is not the sort of person who is willing to passively accept the trust that Bill (or, indeed, anyone else) places in her? In this case, if Bill were to attempt to place his trust in Jane, then it is reasonable to assume that Jane would respond by rejecting his trust. Jane obviously has no interest in fulfilling Bill's trust, but this does not mean she is indifferent to the presence of misplaced trust. Jane may, for example, be concerned about the impact of Bill's trust on her reputation as a trustworthy individual. If Bill trusts Jane, but Jane then fails to fulfill this trust, then Bill may communicate to others that Jane is not the sort of person that can be trusted. If, however, Jane signals to Bill that his trust is misplaced, then she has, in effect, thwarted his attempt to place his trust in her. This is important, for if Jane acts so as to prevent the placement of trust, then there is no sense in which she could be accused of failing to fulfill Bill's trust. One cannot be accused of betraying trust that was never placed, any more than one can be accused of breaking a promise that was never made. If Jane tells Bill that she cannot do what Bill expects her to do, then it is no fault of hers if Bill should persist with his misplaced trust. At the very least, Bill has no right to feel betrayed or disappointed when Jane fails to do what he "trusted" her to do.

Does the mere rejection of Bill's trust mean that Jane is no longer trustworthy relative to Bill? It is hard to see why this should be the case. Suppose someone asks you to do something that you lack the ability to do. You refuse, pointing out, perhaps, that you are simply incapable of doing what the other party expects you to do. Your trustworthiness, in this situation, is not impugned, for you have taken steps to limit the placement of trust to those situations where you are in position to do what you are trusted to do. You cannot be accused of being untrustworthy simply because you have rejected the opportunity to manifest your trustworthiness. What is arguably more injurious to your trustworthiness is to agree to do things that you know you cannot do. In fact, your agreement to do such things is arguably more indicative of your *untrustworthiness* than it is your trustworthiness, for by agreeing to do something that cannot be done, you have allowed someone to trust you in situations where such trust cannot be fulfilled.

At this point, the arguments marshaled in support of two-place trustworthiness start to look a little less convincing. The problem with such arguments is that they overlook the role of trustees in shaping the conditions under which their trustworthiness might be called into question. At first glance, the Bill–Sarah–Jane case in Figure 1b seems impossible to reconcile with an absolutist view of trustworthiness. But this irreconcilability is more illusory than real; it stems from an impoverished conception of the way that trustees can create, cultivate, and change the expectations of those who are in a position to trust them.

A consideration of the active role played by trustees is also useful in formulating a response to three-place trustworthiness. Once again, the arguments in favor of three-place trustworthiness look to be eminently plausible, but much of this plausibility hangs on the idea that trustees are the passive recipients of trust-related actions/evaluations, as opposed to *active* agents that seek to influence these actions/evaluations. To help us see this, let us revisit the case of Robbins' (2016) wife (see Section 2). Robbins suggests that he can trust his wife in respect of certain matters, but not other matters. When it comes to flying a plane, for example, Robbins suggests that his wife ought not to be seen as particularly trustworthy. The upshot is that Robbins' wife cannot be trustworthy in an absolute sense; she must be trustworthy in a more relativistic sense—her trustworthiness is limited to particular spheres of activity or domains of interaction.

The problem here is that Robbins is gauging his wife's trustworthiness relative to a state-of-affairs that has no practical bearing on her actual trustworthiness. Suppose that Robbins were to ask his wife to fly a plane, even though she lacked the ability to fly a plane. What would Robbins' wife say in response to this? Presumably, she would respond by saying "no," and this is perfectly consistent with her status as a trustworthy individual. For why would Robbins' wife acquiesce to her husband's request, if she thought that by doing so she would put her husband's life in jeopardy? To be sure, if Robbins asked his wife to fly a plane believing that she was capable of flying a plane, and his wife then agreed to fly the plane while knowing that she had no flying-related abilities, then she would not qualify as trustworthy. This, however, is not the sort of situation that Robbins is imagining. He is not imagining a situation replete with deception and malign manipulation. Rather, he is assessing his wife's trustworthiness relative to her ability to fly a plane. In short, he is asking himself: "If my wife were to fly a plane, would I trust her?" The answer to this question is, of course, "no." But arguably something important has been omitted from the imaginary exercise. The missing ingredient relates to whether or not Robbins' wife would actually permit her husband to place his trust in her in situations where she would be unable to fulfill that trust. Inasmuch as the answer is "no," then her trustworthiness remains intact. What Robbins' wife is doing here is preventing the possibility of misplaced trust. This, we suggest, is not so much evidence of the absence of trustworthiness as it is the presence of trustworthiness!

At this point, the prospects for an absolutist conception of trustworthiness start to look a little brighter. Inasmuch as we allow for the idea that trustees can act so as to forestall attempts at misplaced trust, then it seems perfectly possible for a trustee to count as trustworthy in an absolutist (or one-place) sense. Such a trustee would be one that could be relied on to do what we trusted them to do, but they would also take steps to prevent us from being led astray. If such a trustee were asked to do something that they could not do (for whatever reason), then they would alert us to this fact, thereby minimizing the chances of their trustworthiness being called into question. At a general level, we might characterize these efforts as the trustee's attempt to minimize expectation-related errors in trust-related situations. The expectations that matter here are those that stem from the trustor's effort to place their trust in the trustee.[7] If X places their trust in Y, then it seems reasonable to think that X must have some sort of expectation regarding Y's subsequent behavior. The reason for thinking this is that if X did not have this expectation, then it is unclear why they would make themselves vulnerable to Y in the first place.[8] At a minimum, then, trust (or, more accurately, the placement of trust) appears to involve an expectation relating to some aspect of the trustee's subsequent behavior. The responses of the trustee can then be understood as an attempt to minimize the error associated with these expectations: Either the trustee will fulfill the trust that is placed in them (in which case they do as they are expected to do), or they take steps to prevent the placement of trust by stating that they cannot do what they are trusted to do. Either way, the trustee is minimizing the possibility of expectation-related errors. By fulfilling trust, the trustee is changing reality so as to conform to the trustor's expectation, while, by rejecting trust, the trustee is changing expectations so as to bring them into alignment with reality. Either way, expectation-related errors are kept to a minimum.

The upshot is what we will call an expectation-oriented account of trustworthiness. Such an account depicts trustworthy entities as those who are concerned with the minimization of expectation-related errors in situations of social dependency (i.e., trust situations). One of the virtues of this account is that it allows for the possibility of an absolutist approach to trustworthiness. What it means to be trustworthy, according to the expectation-oriented account, is to be an entity that is disposed to minimize expectation-related errors in a highly generalized fashion, that is, in a manner where all expectation-related errors have the same (equal) significance. This does not mean that all trustors will be treated the same by the trustee, for the expectations of some trustors will be fulfilled, while the expectations of others will be modified.

---

[7] Note that the expectation-oriented account does not mandate that trustees are required to act in accordance with a potential trustor's expectations outside of a particular trust situation. Consider, for example, a state-of-affairs in which X expects Y to be bad at math simply because Y is a member of a certain social category. The mere presence of this expectation has no bearing on Y's trustworthy status. If Y should turn out to be rather good at math, for example, this will not make Y untrustworthy. Nor does Y need to feign ineptitude at mathematics in order to qualify as trustworthy. The expectations that matter here are those that stem from X's attempt to *place* their trust in Y. Thus, if X trusts Y to fail a math exam, then Y can either reject the offer of trust or seek to challenge X's factually incorrect (and morally dubious) expectations. The point, here, is that one is not trustworthy (or untrustworthy) simply because one acts (or fails to act) in accordance with another's expectations. What matters is the way one avoids the violation of expectations that are established as a result of the trustor's attempt to place their trust in a trustee.

[8] As has been noted by a number of theorists, trust situations typically involve some degree of risk or vulnerability (see PytlikZillig and Kimbrough, 2016)—by placing one's trust in a trustee, a trustor exposes themselves to the risk that the trustor will fail to do what they are trusted to do.

What matters for trustworthiness, under the expectation-oriented account, is not so much the parity of trustors, it is the parity of expectation-related errors that counts.

A second virtue of the expectation-oriented account is that it is readily applicable to nonhuman entities, such as technological systems. This is not to say that all trustees are equally well-equipped to meet the demands of the expectation-oriented account; but there is at least nothing about the account that would make it inadmissible to the likes of, say, a NIS. All that matters for the expectation-oriented account is a disposition to minimize expectation-related errors in trust-related situations. This is arguably something that can be accomplished by both human and technological trustees. The expectation-oriented account is thus just as applicable to AI systems and NISs as it is to human individuals.

A third virtue of the expectation-oriented account concerns the prospects for theoretical integration. In particular, the appeal to expectations (and the minimization of expectation-related errors) is one that is potentially applicable to a number of different accounts of trustworthiness. Consider, for example, that in the earlier discussion of promissory relations, we talked of the promisor acquiring a commitment or obligation to fulfill one's promises. This establishes a point of contact with two recent accounts of trustworthiness, namely, the commitment account (Hawley, 2019) (which ties trustworthiness to the avoidance of unfulfilled commitments) and the obligation-based account (Kelp and Simion, 2020) (which ties trustworthiness to the fulfillment of obligations). At first sight, neither of these accounts seems to have much in common with the expectation-oriented account, but this does not mean that expectations play no role in these accounts. In accepting an offer of trust, for example, Y has plausibly acquired a commitment or obligation to fulfill that trust, but this acceptance is also the basis for an expectation that Y will do what they are trusted to do. And if Y is incapable of doing what they are trusted to do, then they ought to take steps to avoid being bound by these expectations. (Either that, or they should seek to modify the expectations that others have about them.) The point here is that the expectation-oriented account provides us with a particular way of understanding trustworthiness, but this does not mean it is incompatible with other accounts of trustworthiness. In the next section, we will encounter some support for this idea, especially in relation to the commitment account.[9]

## 4. Implications for NISs

Our aim thus far has been to challenge the idea that trustworthiness *must* be conceptualized along relativistic lines. In particular, we have sought to defend the possibility of an absolutist (or one-place) account of trustworthiness—an account that highlights the role of trustees in the minimization of expectation-related errors. It is now time to consider some of the implications of this account for the design, development, and management of NISs. What bearing does the foregoing account have on the practical steps to implement trustworthy NISs?

The first thing to say here is that the risk of expectation-related errors is always apt to be elevated in situations where trustors lack an adequate understanding of the capabilities and limitations of a NIS. If X expects Y to $\Phi$ in situations where Y cannot $\Phi$, then any trust that X places in Y is likely to be misplaced: X will trust Y to $\Phi$, but Y will be unable to fulfill this trust. As noted above, this state-of-affairs can be avoided if Y intervenes in the trust-placement effort. This, however, imposes a considerable burden on Y. In particular, it requires Y to recognize situations in which trust is being placed, to evaluate the feasibility and permissibility of certain actions relative to the contexts in which those actions are to be performed, and to communicate with X in situations where trust cannot be fulfilled. All this is suggestive of a degree of cognitive sophistication and communicative competence that many extant NISs will no doubt lack. There is, however, an alternative strategy here. Note that one of the reasons why misplaced trust occurs is due to a mismatch between what the trustor expects a trustee to do and what the trustee can actually do. Perhaps, for example, X believes Y has certain abilities when in fact they do not. The result is that X places their trust in Y, expecting Y to $\Phi$, and X is subsequently surprised (and not in a good way!) when $\Phi$ fails to materialize. This situation could be remedied if X had a better understanding of Y. In the extreme case, if X knew everything there was to know about Y, then X would already know whether or not their trust was likely to be fulfilled in a given situation. Accordingly, X would refrain from placing their trust in Y in those situations where Y was incapable of fulfilling that trust. In this situation, then, there would be no reason for Y to intervene in the trust placement effort.

---

[9] It is worth noting (although only as a sort of coda to the main story) that the expectation-oriented account establishes a point of contact with so-called predictive processing and/or free energy minimization approaches in cognitive science (Friston, 2010; Clark, 2016). This speaks to issues of theoretical integration that reach beyond the realm of the trust-related literature. See Smart (2021), for more on this.

It is, of course, unrealistic to expect stakeholders (including users) to know or understand *everything* there is to know about a NIS. Nevertheless, the foregoing discussion highlights the importance of initiatives that are directed at shaping stakeholder expectations. The aim, here, is not to persuade stakeholders that a particular NIS is trustworthy; it is ensure that stakeholders do not have unrealistic or inaccurate expectations of a NIS. The goal, in short, is to ensure that stakeholders possess an accurate understanding of what the NIS can do, and perhaps what they need to do to ensure the NIS operates as expected.

It is here that we encounter a potentially interesting link with a recent account of trustworthiness that is attributable to Sung (2020). According to this account, the Confucian notion of *xin* resembles what we commonly (in the West) refer to as trustworthiness. One who is *xin*, Sung suggests, attempts to ensure that others perceive one according to the way one actually is. Sung contrasts her account of trustworthiness with those that focus on the trustee's responses subsequent to the placement to trust. Thus, rather than focus on the role of the trustee in fulfilling trust, the Confucian account draws attention to the way in which trustees can provide the basis for mutually beneficial trust relations by engaging in sincere self-presentational activities—activities that ensure others' have an accurate understanding of the sort of person one is. In one sense, this does mark a genuine departure from the main thrust of theoretical debates pertaining to trustworthiness. But to see the Confucian account as somehow separate or distinct from other accounts of trustworthiness is, we suggest, a mistake. From the standpoint of an expectation-oriented account, there is no real incompatibility between the effort to ensure that others perceive oneself according to the way one is and the effort to respond to the placement of trust by fulfilling trust. Both these efforts work to ensure that expectation-related errors are kept to a minimum; it is just that the Confucian account assigns greater emphasis to the role of trustees in influencing trustor expectations prior to the placement of trust, thereby helping to reduce the chances of misplaced trust. In short: If X has access to reliable information about Y (via Y's sincere self-presentational activities), then X is in a position to form accurate expectations about Y, and this helps to minimize the sorts of confusions and misunderstandings that can culminate in misplaced trust.

A consideration of trustor expectations thus helps to highlight one of the ways in which an existing theoretical account of trustworthiness (the Confucian account) might be applicable to NIS. Another point of contact arises in respect of the commitment account of trustworthiness (Hawley, 2019). According to the commitment account, trustworthiness is best understood in terms of commitment: "[T]o be trustworthy," Hawley (2019, p. 23) suggests, "is to live up to one's commitments, while to be untrustworthy is to fail to live up to one's commitments." Thus conceived, trustworthiness is about the avoidance of unfulfilled commitments. This requires caution in incurring new commitments, as well as diligence in fulfilling commitments that have already been acquired.

From the standpoint of the expectation-oriented account, a commitment serves as the basis for expectations regarding trustee behavior. Accordingly, if a trustee is trustworthy, then the trustee can be expected to behave in accordance with their commitments. In a NIS context, such commitments might be embodied in publicly communicated policy declarations—explicit, human-readable statements of what stakeholders can expect from a NIS. Examples of such commitments might include those relating to data privacy, compliance with ethical standards, and a respect for human rights (see Beduschi, 2021).

But commitments need not be intended solely for human consumption. Given that trustworthy NISs are required to coordinate their behavior with regard to a set of commitments (each of which embodies an expectation), there may be situations in which the permissibility of certain actions needs to be evaluated as part of the routine operation of a NIS. Imagine, for example, a situation in which a NIS is bound by a privacy-related commitment and a government official then requests access to information in a way that violates this commitment. Inasmuch as the NIS is trustworthy, then it should act in accordance with its commitments, which means that the information access request must be rejected. In order to do this, however, it would clearly be useful to have a means of assessing or evaluating the permissibility of potential actions—of determining whether or not a given action (in this case, an information access request) is compatible with preexisting commitments. It is here that we encounter a potentially fruitful link with work into policy-aware computing (Weitzner et al., 2006), purpose-based data access protocols (Kraska et al., 2019), and regulatory compliance modeling (Taylor et al., 2021). Work in these areas seeks to ensure that the functional operation of a system is consistent with a set of constraints, many of which are tied to human-readable policy statements. There is, as far as we can tell, no reason why these constraints could not be construed as commitments, albeit ones that are rendered in machine-readable form.

Commitments are important, for they provide a means by which the trustworthiness or untrustworthiness of a trustee can be assessed. Just as one is not a good promisor if one routinely breaks one's promises, then one cannot be a "good" (i.e., trustworthy) trustee if one routinely fails to live up to one's commitments. There are, however, situations in which a trustee can be put in an impossible position. Consider a situation in which two military forces (blue and red) are engaged in an armed conflict. A blue force soldier is, let us suppose, committed to always act in the interests of blue force. Suppose, however, that part of what it means to act in the interests of blue force is to solicit the trust of red force and then betray that trust. In this situation, the blue force soldier has no option but to violate the trust of one or other party: either the soldier must betray the trust of blue force (e.g., by becoming a double agent) or they must fulfill the trust of blue force, thereby betraying the trust of red force.

In a NIS context, such situations can arise if a single stakeholder has an inordinate amount of influence over a NIS. If, for example, a NIS is committed to always do what one particular stakeholder instructs it to do, then the door is open to situations in which the relativity of trustworthiness is assured. If, for example, a NIS must always do what a government official requests it to do, and the government official then requests access to information that violates a privacy-related commitment, then the NIS has no option but to act in an untrustworthy manner: it must either violate the trust of those who assume that their information will be protected, or it must violate the trust of those who assume that the NIS will always do what they want it to do.

The dilemma is real, but this does not mean it cannot be avoided. What the dilemma reveals is the importance of ensuring that commitments are formulated in such a way as to prevent the possibility of competing commitments. In formulating an initial set of commitments, designers, developers, and policy-makers are obliged to give due care and consideration to what is entailed by such commitments. Commitments relating to data privacy, security, unauthorized disclosure, and so on, are fine by themselves, but they mean nothing if the presence of another commitment requires the NIS to renege on these commitments.

This is not to say that commitments ought to be regarded as universally applicable "rules" that cannot be overridden in certain circumstances. A privacy-related commitment, for example, may only apply in situations where an individual citizen is a member of the class of law-abiding citizens. Such a commitment would not apply, if, for example, the individual was known to have perpetrated a terrorist attack. In this case, we would naturally expect the privacy-related commitment to be rescinded. By itself, this need not mean that a NIS is untrustworthy, for this would only be the case if the NIS had made a prior commitment to act in the interests of *all* individuals no matter their social status. Commitments, however, are typically contingent on certain states-of-affairs. An employee may acquire a commitment as a result of working for an organization, but they need not be beholden to that commitment if the circumstances should change. If they are fired, for example, their social status will change, and they will no longer qualify as an employee. In this case, it would make no sense to say that they are still beholden to the commitments that were acquired during the course of their former employment.

Trustor expectations can be informed by one's previous experience of interacting with a trustee. They can also be informed by whatever information one has about a trustee. These, however, are not the only ways that trustor expectations can be cultivated. Just as trustees can play an active role in shaping trustor expectations, trustors can, on occasion, play an active role in ensuring that their own expectations are suitably aligned with reality. Suppose, for the sake of example that one wants to ensure that a newly purchased smartphone operates in accordance with one's expectations. One way (and perhaps the only way!) of accomplishing this is to configure the device in some way, either by changing the device's default settings, or by removing/installing apps. Perhaps, then, we can see a role for trustors in establishing the trustworthiness of a trustee. Inasmuch as trustworthiness is tied to the minimization of expectation-related errors, then perhaps trustworthiness is as much to do with the trustor exercising control over the trustee as it is with the trustee responding to the placement of trust in an appropriate manner.

In a NIS context, this idea introduces us to capabilities that are intended to give individual trustees greater control over identity-related information. Such capabilities typically form the basis of proposals relating to so-called user-centric or self-sovereign identity systems (see Tobin and Reed, 2017; Domingo and Enríquez, 2018). Examples include individualized forms of consent, control, and authorization that limit the access of relying parties to certain types of information; a degree of discretion regarding what a system will commit itself to doing in certain situations; and the implementation of safeguards that prevent the possibility of inadvertent harm to the trustor. Insofar as such measures help to improve the predictability of a NIS, then there is, we suggest, no reason why they cannot be seen to play a productive role in enhancing the NIS's trustworthiness.

We have now surveyed some of the ways that an expectation-oriented account of trustworthiness might inform the design, development, and management of NISs. One of the issues to arise from this discussion relates to the role of intelligence in securing trustworthiness. To be trustworthy, one needs to ensure that one does not accept an offer of trust that one is ill-disposed to fulfill. But this sort of evaluative effort comes with an attendant set of cognitive and communicative demands. Some insight into these demands is revealed by a consideration of human–human trust relationships. If someone asks you to $\Phi$, for example, then you need to evaluate whether you are able to $\Phi$ in circumstances where $\Phi$ is expected to occur. This requires an understanding of your own abilities, as well as (perhaps) your access to certain resources. It also requires you to consider the request relative to your preexisting commitments, so as to avoid the possibility of competing commitments. If you cannot agree to do what you are asked to do, then you need to communicate this to your interlocutor. You may also be required to provide an explanation as to why you are rejecting an opportunity to manifest your trustworthiness on this particular occasion.

All this serves as a potent reminder of the importance of cognitive and communicative abilities to human trust relationships. This, however, raises an important question: Inasmuch as our own (human-level) cognitive/communicative abilities play a crucial (and perhaps indispensable) role in ensuring our own trustworthiness, then is there any reason to think that entities bereft of these abilities will be fitting targets for trust-related evaluations? The answer to this question remains unclear, although the continued disparity between human and machine intelligence may be one of the reasons why some theorists are reluctant to apply trust-related terms to nonhuman entities (e.g., AI systems) (see Ryan, 2020). Notwithstanding this worry, we hope to have shown that a consideration of cognitive and communicative abilities is likely to be important when it comes to the development of trustworthy NISs. While AI systems are typically seen to inspire their own trust-related concerns (e.g., Andras et al., 2018), the present analysis suggests that the integration of AI capabilities into NISs is likely to be an important focus area for future research.

## 5. Conclusion

A trustworthy NIS is one that ought to be trustworthy to all actual and potential trustors. But given the scale of the trustor communities associated with NISs, it is unclear whether a single NIS can be equally trustworthy to all trustors, especially when those trustors have competing or conflicting interests. This worry is accentuated by the idea that trustworthiness ought to be conceptualized along relativistic lines—a position that is widely endorsed by theoretical accounts of trustworthiness (Hardin, 2002; Jones, 2012).

In the present article, we outlined an approach to trustworthiness dubbed the expectation-oriented account. According to this account, trustworthiness corresponds to a disposition to minimize expectation-related errors in situations of social dependency (i.e., trust situations). Such a disposition can be manifest in a variety of ways. One way it can be manifest is by the trustee's attempt to fulfill the trust that is placed in them; a second way is for the trustee to intervene in the trust placement effort so as to preclude the possibility of misplaced trust; and a third way is by engaging in open, honest, and transparent communication about the sort of entity the trustee is (e.g., the sorts of abilities the trustee possesses).

These features, we suggest, are perfectly compatible with *both* a relativistic and an absolutist approach to trustworthiness. What marks the distinction between these approaches is the value assigned to expectation-related errors: Trustworthiness can be relative, in the sense that some trustor expectations are assigned greater significance than others; or it can be absolute, in the sense that all trustor expectations are assigned equal significance. For NISs, we suggest that it is the latter, absolute form of trustworthiness that best serves the interests of the scientific, engineering, and policy-making communities.

**Acknowledgment.** A preprint of this article is available via arXiv: https://doi.org/10.48550/arXiv.2112.09674.

**Funding Statement.** This work was supported, in whole or in part, by the Bill & Melinda Gates Foundation (INV-001309). Under the grant conditions of the Foundation, a Creative Commons Attribution 4.0 Generic License has already been assigned to the Author Accepted Manuscript version that might arise from this submission.

**Competing Interests.** The authors declare no competing interests exist.

**Author Contributions.** Conceptualization: P.R.S., W.H.; Writing—original draft: P.R.S.; Writing—review and editing: P.R.S., W.H., M.B. All authors approved the final submitted draft.

**Data Availability Statement.** Data availability is not applicable to this article as no new data were created or analyzed in this study.

**Ethical Standards.** The research meets all ethical guidelines, including adherence to the legal requirements of the study country.